\documentstyle[12pt,epsf]{article}
\newcommand{\be}{\begin{equation}}
\newcommand{\ee}{\end{equation}}
\newcommand{\bea}{\begin{eqnarray}}
\newcommand{\eea}{\end{eqnarray}}
\newcommand{\brr}{\begin{array}}
\newcommand{\err}{\end{array}}
\newcommand{\bc}{\begin{center}}
\newcommand{\ec}{\end{center}}

\newcommand{\mlmsbar}{m^{\overline{\rm MS}}_\ell}
\newcommand{\msmsbar}{m^{\overline{\rm MS}}_s}

\newcommand{\mev}{\,{\rm MeV}}   
\newcommand{\gev}{\,{\rm GeV}}   

\newcommand{\epse}{\varepsilon^{\prime}/\varepsilon}

\newcommand{\asz}{\alpha_s(M_Z)}
\newcommand{\msms}{m_s^{\overline{MS}}}

\newcommand{\prd}{Phys.~Rev. }
\newcommand{\pl}{Phys.~Lett. }
\newcommand{\np}{Nucl.~Phys. }
\def\ee{\end{equation}}

\newcommand{\ba}{\begin{eqnarray}}  
\newcommand{\ea}{\end{eqnarray}}

\newcommand{\as}{\alpha_s} 
\newcommand{\J}{\hat{J}}
\newcommand{\W}{\hat{   W}}
\newcommand{\U}{\hat{ U}}
\newcommand{\gammaz}{\hat{\gamma}^{(0)T}}

\begin{document}
\pagestyle{empty} 
\begin{flushright}
ROME  99/1268\\
% RM3-TH/99-xxx
\end{flushright}
\vskip 1cm
\centerline{\LARGE{\bf{$\epse$ from Lattice QCD}}}
\vskip 1cm
\centerline{\bf{M. Ciuchini$^{a}$, E. Franco$^{b}$, L. Giusti$^{c}$,
V. Lubicz$^{a}$, G. Martinelli$^{b,\dagger}$}}
\vskip 0.3cm
{\small
\centerline{$^a$ Dipartimento di Fisica, Universit\`a di Roma Tre
and INFN, Sezione di Roma Tre}
\centerline{Via della Vasca Navale 84, I-00146 Rome, Italy}
\centerline{$^b$ Dipartimento di Fisica, Universit\`a di Roma ``La Sapienza''
and INFN, Sezione di Roma,}
\centerline{P.le A. Moro 2, I-00185 Rome, Italy}
\centerline{$^c$ Department of Physics, Boston University, Boston, MA 02215 
USA.}
}
\begin{abstract}
Lattice calculations of matrix elements relevant for kaon decays,
and in particular for $\epse$, are reviewed. The r\^ole of the strange quark mass
is also discussed. A comparison  with other non-perturbative approaches 
used to compute kaon decay amplitudes is made.
\end{abstract}
\vfill
\begin{flushleft}
$\dagger$ Presented by G.~Martinelli.
\end{flushleft}
\eject
\pagestyle{empty}\clearpage
\setcounter{page}{1}
\pagestyle{plain}
\newpage
\pagestyle{plain} \setcounter{page}{1}
\section{Introduction}
\label{sec:intro}
Theoretical predictions for non-leptonic decays are obtained by  introducing
an effective low-energy Hamiltonian expressed in terms of local operators and of
the corresponding Wilson coefficients. The latter can be computed in
perturbation theory, whereas the matrix elements of 
the  operators have to be evaluated within some non-perturbative approach.  
For kaon  decays, the Wilson coefficients are known at the 
next-to-leading order accuracy~\cite{alta}--\cite{ciuc2} and the main uncertainties come
from the calculation of the matrix elements. 
In this talk we review
the present status of lattice computations of matrix elements which are relevant
in kaon decays.  We focus, in particular, on those which enter  the
calculation of $\epse$ and make also a comparison with other 
methods, namely the Chiral Quark Model ($\chi$QM) and the 
large $N$ expansion~\cite{bert1}--\cite{paschoslast}. 
For a more general discussion of the theory of CP violation
in kaon decays see~\cite{burask99} and references therein. 
\par  Some general remarks are necessary before entering a more detailed
discussion. Given the large
numerical cancellations which may occur in the theoretical expression of $\epse$,
a solid prediction should avoid the ``Harlequin procedure". This procedure
consists in patching together $B_6$ from the $\chi$QM, $B_8$ from the
$1/N$ expansion, $\msms$ from the lattice, etc., or any other
combination/average of different  methods.  All these quantities 
are indeed strongly correlated (for example $B_6$ and
$B_8$ in the $1/N$ expansion or  $B$ parameters and  quark masses
in the lattice approach~\cite{giusti}) and should be consistently computed
within  each given theoretical framework.
 Unfortunately,  none  of the actual non-perturbative
methods is in the position to avoid completely the Harlequin procedure,
not even for the most important input  parameters only.
The second important issue is the consistency of the renormalization
procedure adopted in the perturbative calculation of the Wilson
coefficients and in the non-perturbative computation  of  the
operator matrix elements. This problem is particularly serious for the
$\chi$QM and the $1/N$ expansion, and will be discussed when comparing the
lattice approach to these methods. We will address, in particular, the 
problem of the quadratic divergences appearing in the $1/N$ expansion.
This is an important issue, since the authors of
ref.~\cite{ham,paschoslast} find that these divergences 
provide the enhancement necessary to explain the large values of $Re
A_0$ and of $\epse$.    
\par Schematically, $\varepsilon^{\prime}$ can be cast in
the form \be \varepsilon^{\prime} = \frac{\exp (i\pi/4 )}{\sqrt{2}} \frac 
{\omega}{ReA_{0}}\times \left[ \omega^{-1} ImA_{2} - (1-\Omega_{IB}) ImA_{0} \right] \ee
where $\omega=ReA_{2}/ReA_{0}$  and $ReA_{0}$ are taken from  experiments,  and 
$\Omega_{IB}$ is a correcting factor, estimated in
refs.~\cite{don}--\cite{lu}, due 
to isospin-breaking effects.  Using the operator product expansion, 
the  $K \to \pi \pi$ amplitudes $ImA_{2}$ and $ImA_{0}$  are computed from the 
matrix elements of the effective Hamiltonian, expressed in terms of 
Wilson coefficients and renormalized operators 
\be \langle\pi\pi\vert {\cal H}^{\Delta S=1}\vert K^0\rangle_{I=0,2}=
\sum_{i}\langle \pi\pi\vert Q_{i}(\mu) \vert
K^0\rangle_{I=0,2} \,   C_{i}(\mu) \, \ee
where the sum is over a complete set of operators, which depend on 
the renormalization scale $\mu$.  
Wilson coefficients and   matrix elements of the  operators $Q_{i}(\mu)$, 
appearing in the effective Hamiltonian, separately
depend on the choice of the renormalization scale and scheme. This dependence  cancels in
physical quantities, such as $ImA_{2}$ and $ImA_{0}$, up to higher-order corrections in the perturbative expansion of
the Wilson coefficients. For this crucial cancellation to take  place, the non-perturbative method 
used to compute hadronic matrix elements must allow    a definition of the
renormalized operators consistent with the scheme used in the 
calculation of the Wilson coefficients. 
\par So far, lattice QCD is the only 
non-perturbative approach in which both the scale and 
scheme dependence can be consistently accounted for,
using either lattice perturbation theory or non-perturbative renormalization 
techniques~\cite{npm,npm4f}. This is the main reason why the authors of
refs.~\cite{ciuc2,ciuc1,ciuc3,newnoi} have followed  this approach over the years. 
\par There is a general consensus~\cite{burask99} that the largest contributions are those 
coming from $Q_{6}$ and $Q_{8}$ (for $ImA_{2}$), with opposite sign, and sizeable 
contributions may come from $Q_{3}$, $Q_{4}$  and $Q_{9}$ (for 
$ImA_{2}$) in the presence of large cancellations
between  $Q_{6}$ and $Q_{8}$, i.e.  when the prediction for $\epse$  
is of  ${\cal O}(10^{-4})$~\footnote{The
operator $Q_{3,4,5,6}$ only contribute to the $I=0$ amplitudes.}. For this
reason the following discussion, and the comparison with other 
calculations, will be focused on the determination, and errors, of the 
matrix elements of  the two most important operators.
\section{Matrix elements from lattice 
QCD}
\label{sec:me}
The evaluation of physical $K \to \pi\pi$ matrix elements on the lattice relies
 on the use of Chiral
Perturbation Theory ($\chi$PT): so far only $\langle \pi \vert Q_{i}(\mu) \vert
 K \rangle$ and  $\langle 
\pi(\vec p=0) \pi(\vec q=0) \vert Q_{i}(\mu) \vert K \rangle_{I=2}$  
(with the two pions at rest) have been computed for a variety of operators. The
 physical matrix elements 
are then obtained by using  $\chi$PT at the lowest order. 
This is a consequence of the difficulties in extracting physical multi-particle
 amplitudes  in Euclidean space-time~\cite{mt}. Proposals to overcome this 
problem have been presented, at the price of 
introducing some model dependence in the lattice results~\cite{cfms}. 
The use of $\chi$PT implies that  large systematic  errors may occur in the 
presence of large   corrections   from higher-order terms  in the chiral
expansion and/or from FSI. 
This problem is common to all approaches: if large higher-order terms
 in the chiral expansion are indeed present and 
important,  any   method aiming to have these 
systematic errors under control must be able  to reproduce 
the FSI phases of the physical amplitudes. The approaches of 
ref.~\cite{bert1,bert2} and \cite{1oN2,ham,paschoslast,1oN}, however, give FSI
smaller than their physical values.
\par 
\underline{$I=2$ matrix element of $Q_8$.} There exists a large set
of quenched calculations of $\langle Q_{8} \rangle_{2}$ 
performed with different formulations of the lattice fermion actions 
(Staggered, Wilson, tree-level  improved, tadpole improved) and renormalization
techniques (perturbative, boosted perturbative, non-perturbative), at
several values of the inverse lattice spacing $a^{-1}=2\div 3
$ GeV~\cite{giusti,npm4f,kgs,bgs,lellolin}. 
All these calculations, usually expressed in terms of 
$B_{8}^{(3/2)}$,  give consistent results within $20 \%$ of uncertainty.
Among the results, in the numerical estimates presented in sect.~\ref{sec:numerical},
we have taken the central value from the recent 
calculation of ref.~\cite{giusti}, 
where the matrix elements  $\langle Q_{8} \rangle_{2}$ and 
$\langle Q_{7} \rangle_{2}$ have been 
computed without any reference to the quark masses, and inflated the 
errors to account from the uncertainty due to the quenched 
approximation (unquenched results are expected very soon) and the lack 
of extrapolation to zero lattice spacing. 
For a discussion on the r\^ole of quark masses  see below at the end of this section.
\par \underline{Matrix element of $Q_6$.} For $\langle Q_{6} \rangle_0$
from the lattice, the situation 
appears worse today than a few years ago when the calculations of 
refs.~\cite{ciuc2,ciuc1,ciuc3} were performed: \par
 i) until 1997, the only existing lattice
result, obtained with staggered fermions (SF) without NLO lattice perturbative
corrections,  was  $B_{6}=1.0 \pm 0.2$~\cite{kilcup}. This is the value
used in  previous analyses~\cite{ciuc2,ciuc1,ciuc3};
\par ii) with SF even more accurate results have been quoted recently, namely 
$B_{6}=0.67\pm 0.04\pm 0.05$ (quenched)  and  $B_{6}=0.76\pm 0.03\pm 
0.05$ (with $n_{f}=2$)~\cite{k2};  \par iii) ${\cal O}(\alpha_{s})$ 
corrections, necessary to match lattice operators to  continuum
ones at the NLO, are so huge for $Q_6$ in the case of SF 
(in the neighbourhood of $-100 \%
$~\cite{stagpert}) as to make   all the above results unreliable.  Note,
however, that the corrections tend to diminish the value of $\langle
Q_6\rangle_0$; 
\par iv) the latest lattice results for this matrix element, computed
with  domain-wall fermions~\cite{newsoni} from $\langle \pi \vert Q_6\vert K\rangle$, 
are absolutely surprising: $\langle Q_6\rangle_0$ has a sign opposite to what expected 
in the VSA,  and to what is found with the $\chi$QM and the $1/N$ expansion. 
Moreover, the absolute value is so large as to give $\epse \sim - 120 \times 10^{-4}$. 
Were this confirmed,  even the conservative statement by  Andrzej Buras~\cite{burask99}, namely 
{\it ... that 
certain features present in the Standard Model are confirmed by the experimental results. 
Indeed
the sign and the order of magnitude of $\epse$ predicted by the SM turn out to agree with the
data...} would result too optimistic.
In order to reproduce the experimental number, $\epse \sim 20 \times 10^{-4}$, not only new
physics is required, but a large cancellation should also occur between the Standard Model
and  the new physics contributions.   
Since this result has been obtained with domain-wall fermions, a lattice formulation
for which numerical studies started very recently, and no details  on 
the renormalization and subtraction procedure have been given, we consider 
premature to use the value of the matrix element of ref.~\cite{newsoni} in phenomenological
analyses. Hopefully, new lattice calculations will clarify this fundamental issue.
\par \underline{$B$ parameters and quark masses.}
Following the common lore,  matrix elements of  weak four-fermion operators are 
given in terms  of the  so-called $B$-parameters which measure 
the deviation  of their values from those obtained in the Vacuum 
Saturation Approximation (VSA).  A classical example is provided by the 
matrix element of the $\Delta S=2$ left-left operator  $Q^{\Delta S=2}=
 \bar s \gamma_\mu (1- \gamma_{5} ) d  \; \bar s \gamma^\mu (1- \gamma_{5} ) 
 d$  relevant to the prediction of   the CP-violation parameter $\varepsilon$
\be  \langle  \bar K^{0} \vert   Q^{\Delta S=2}  \vert K^{0} 
\rangle = \frac{8}{3} M_{K}^{2} f_{K}^{2} B_{K} \ . \ee
VSA values and $B$-parameters are also used for matrix elements of 
operators entering the expression of  $\epse$, in 
particular 
$Q_{6}=\bar s_\alpha\gamma_\mu(1-\gamma_5)d_\beta\sum_q \bar q_\beta 
\gamma_\mu (1+\gamma_5) q_\alpha$ and $Q_{8}= 3/2 
\bar s_\alpha\gamma_\mu(1-\gamma_5)d_\beta\sum_q e_q \bar q_\beta 
\gamma_\mu (1+\gamma_5) q_\alpha$
\bea\label{eq:Bbrutti}
\langle \pi\pi|  Q_6(\mu) | K^{0}\rangle_{I=0} & = & - 4 
\left[\frac{M^2_{K^0}}{m_s(\mu)+ m_d(\mu)}\right]^2 (f_K - f_\pi) 
\; B_6(\mu)\nonumber\\
\langle \pi\pi| Q_8(\mu) |K^0\rangle_{I=2} & = &
\sqrt{2} f_\pi \left[
\left(\frac{ M^2_{K^0} }{ m_{s}(\mu) + m_d(\mu) }\right)^{2}\right.\nonumber\\ 
& -& \left.\frac{1}{6}\left(M_K^2 - M_\pi^2\right)\right] \;
B^{(3/2)}_8(\mu)\; .
\eea
 Since in the VSA and in the $1/N$ 
expansion the expression of the matrix elements 
is quadratic  in $m_{s} + m_d$, predictions for the physical 
amplitudes  are  heavily affected by the specific value which we  assume for 
this quantity. Contrary to $f_{K}$, $M_{K}$, 
quark masses  are not  directly measured by experiments and the present accuracy  
in their determination is still rather  poor~\cite{qmasses,sr}.
Therefore,  the ``conventional'' parametrization 
induces  a large systematic  uncertainty in the prediction of the 
physical amplitudes  of $\langle  Q_6\rangle_{I=0} $ and 
$\langle  Q_8 \rangle_{I=2} $ (and of any other left-right 
operator).  Moreover, whereas for $Q^{\Delta S=2}$ we introduce $\hat B_{K}$ 
as an alias of the matrix element,  by using~(\ref{eq:Bbrutti}) we 
replace each of the matrix elements with 2 unknown quantities, i.e. 
the $B$-parameter and $m_{s} + m_d$.   Finally, in many 
phenomenological analyses, 
the values of the $B$-parameters of $\langle  Q_6\rangle_{I=0}$ and 
$\langle  Q_8 \rangle_{I=2}$ and of the quark masses are taken by independent lattice 
calculations, thus increasing the spread of the theoretical  predictions.
All this can be avoided in the lattice approach, where matrix elements can be computed 
from first principles. 
In ref.~\cite{giusti} a new   parametrisation of the   matrix elements in terms of well known
experimental  quantities, without any reference    the strange (down) quark mass, has been
introduced.  
This results in  a determination of  physical amplitudes with  smaller systematic errors.
The interested reader can refer to \cite{giusti} for details.
\par Before ending this  discussion, we wish  to  illustrate   the
correlation existing between  the $B$ parameters and the
quark masses  in lattice calculations. 
 On the lattice,  quark masses are often extracted 
from the matrix elements of the (renormalized) axial current
($A_\mu$)   and pseudoscalar density ($P(\mu)$) (for simplicity we assume
degenerate quark masses)
\be   m(\mu) \equiv \frac{1}{2} \frac{\langle \alpha \vert \partial_\mu 
A_\mu  \vert \beta \rangle}{\langle \alpha \vert  P(\mu) \vert \beta
\rangle} \, , \label{eq:awi} \ee
where $\alpha$ and $\beta$ are physical states (typically
$\alpha$ is the vacuum state and $\beta$ the one-pseudoscalar meson state)
and $m(\mu)$ and $P(\mu)$ are renormalized in the same scheme.
On the other hand, the $B$ parameters of  $Q_6$ and $Q_8$
are obtained (schematically) from the ratio of the following  
matrix elements, evaluated  using  suitable ratios of
correlation functions~\footnote{ See for example ref.~\cite{npm4f}. 
For simplicity the superscript $(3/2)$ in $B_8$ is omitted.}:
\be B_{6,8}(\mu)  \propto \frac{\langle \pi \vert Q_{6,8}(\mu) \vert K
\rangle}{\langle \pi \vert P_\pi(\mu) \vert 0 \rangle\langle 0 \vert
P_K(\mu) \vert K \rangle} \, , \label{eq:bpar}\ee
where $P_\pi$ and $P_K$ are the   pseudoscalar densities with the flavour 
content of the pion or  kaon, respectively.
Eqs.~(\ref{eq:awi}) and (\ref{eq:bpar}) demonstrate the strong correlation 
existing between $B$ parameters and quark masses: large values of the
matrix elements of $P(\mu)$ correspond, at the same time, to small values
of $m(\mu)$ and  $B_{6,8}(\mu)$.
Physical amplitudes, instead, behave as 
\be \langle Q_{6,8}\rangle = \mbox{const.} \times
\frac{B_{6,8}(\mu)}{m(\mu)^2}
\, , \label{eq:ratio}\ee 
where  ``const." is a constant which may be expressed in terms of
measurable quantities (specifically $M_K$ and $f_K$) only. 
From  eqs.~(\ref{eq:awi}) and (\ref{eq:bpar}),  
we recognize that the dependence on $\langle P(\mu) \rangle$
cancels in the  ratio $B_{6,8}/m(\mu)^2$, appearing in
the physical matrix elements. 
\par Previous lattice studies  preferred to work with 
$B$ parameters because these are dimensionless quantities,  not 
affected by the uncertainty due to the calibration of the lattice spacing.
This method can still be used, provided that quark masses and the $B$
parameters from the same simulation are  presented together 
(alternatively one can give directly the ratio $B_{6,8}/m(\mu)^2$). 
In ref.~\cite{giusti},  two possible definitions of dimensionless
``$B$ parameters", which can be directly related to physical matrix
elements without using the quark masses have been proposed.
\par \underline{The strange quark mass.} Although
in lattice calculations of  matrix elements any reference 
to  quark masses can be avoided,
these are fundamental parameters of the Standard Model and are used 
in the large $N$
expansion.  Here we would like to add only a few remarks to ref.~\cite{sr}, 
where this subject has been reviewed. 
\par The first observation is the following.  Lattice calculations  that use
non-perturbative renormalization methods obtain the quark masses without errors coming from 
the truncation of the perturbative series (typically in the RI-MOM or the 
Schr\"odinger
renormalization schemes; for a complete set of references see ref.~\cite{sr}).  
The conversion of these results to the ``standard" $\overline{MS}$
scheme can be done at the N$^3$LO.  Differences between NLO, N$^2$LO and  N$^3$LO
are important, $\sim 6\div 10$ MeV for $\msmsbar$,  as demonstrated by the following example
\bea
\label{MSmasses}
&& \hspace*{3.2cm}  {\scriptsize{\textsf{NLO}}} \quad  \; \; 
 {\scriptsize{\textsf{N$^{\sf 2}$LO}}}\quad   \; \;  
  {\scriptsize{\textsf{N$^{\sf 3}$LO}}} \quad \cr 
&& \hfill \cr
&& \mlmsbar (2\gev) = \left\{ 5.2(5);\ 4.9(5);\ 4.8(5) \right\}\ \mev
\, ,\cr
&&\cr
&& \msmsbar (2\gev) = \left\{ 120(9);\  114(9);\ 111(9)\right\}\ \mev
\, ,
\eea
taken from ref.~\cite{damir}.
Therefore, when confronting results from different calculations
it is  necessary to specify  the order at which the results have been obtained. 
In table 1 of ref.~\cite{sr},  results obtained with perturbation theory 
at NLO or with non-perturbative
methods at the N$^3$LO  are directly compared.  This, for the reasons discussed above, is
misleading.  We also note that in most of the phenomenological applications, 
for example with QCD sum rules,
the theoretical expressions are only known at the NLO and, for consistency, 
quark masses at the same level of accuracy should be used. 
\par By comparing the results   obtained with the
non-perturbatively improved action  at $\beta=6.2$ (corresponding to $a^{-1} \sim 2.6$ GeV,
which is the value used by the APE Collaboration)
with those extrapolated to the continuum (table~2 of ref.~\cite{wittig}), 
one finds  the discretization errors at
this value of the lattice spacing and with this action to be $3\div 4 \%$. This is much 
smaller than the $15\%$ quoted in \cite{sr}.
\par The large reduction  of the value of the masses in the unquenched case,
found by the CP-PACS Collaboration, is not confirmed by other lattice calculations by the 
MILC~\cite{claude} and APE Collaborations~\cite{gimenez} and it is at variance with the bounds 
of ref.~\cite{taron}.
We think that further investigation is required on this important point.
\section{Renormalization group invariant operators}
\label{sec:rgi}
 Wilson coefficients  and  renormalized operators 
  are usually defined  in a given   scheme ($HV$, $NDR$, 
$RI$), at a fixed renormalization scale $\mu$, and depend on the 
renormalization scheme and scale. This is a source of confusion in 
the literature. Quite often, for example, one finds comparisons of 
$B$ parameters  computed in different schemes. Incidentally, we note that the 
$NDR$ scheme used in the lattice calculation of ref.~\cite{bgs} 
differs from the standard $NDR$ scheme of refs.~\cite{bur2}--\cite{ciuc2}; on the other 
hand, the $HV$ scheme of ref.~\cite{bur2} is not the same as the $HV$ 
scheme of ref.~\cite{noi}. In some cases, the differences between 
different schemes may be numerically large, e.g. $B_{8}^{(3/2) HV}\sim  
1.3 \; B_{8}^{(3/2) NDR}$ at $\mu \sim 2$ GeV.
To avoid all these problems, it is convenient to introduce a Renormalization
Group Invariant (RGI) definition of Wilson coefficients and composite operators 
which generalises what is usually done for $B_K$ using the
RGI $B$-parameter $\hat B_K$.  The idea is very simple. Physical amplitudes can be written as 
\be 
\langle F \vert {\cal H}\vert I\rangle = 
\langle F \vert \vec{Q}(\mu) \vert I \rangle \cdot \vec{C}(\mu) 
\, , \label{wope} 
\ee
where $\vec{ Q}(\mu) \equiv
(  Q_1(\mu),   Q_2(\mu),   \dots,  Q_N(\mu))$  is the operator basis
 and  $\vec{C}(\mu)$  the corresponding Wilson coefficients, represented as a column vector. 
$\vec C(\mu)$ is  expressed in terms of its counter-part, computed at a large scale $M$, 
through the renormalization-group evolution matrix  $\W[\mu,M]$
\be 
\vec C(\mu) = \W[\mu,M] \vec C(M)\, . \label{evo} 
\ee
\par The initial conditions for the evolution equations,
$\vec C(M)$, are obtained by perturbative matching 
of the full theory, which includes
propagating heavy-vector bosons ($W$ and $Z^0$), 
the top quark, SUSY particles, etc., to the effective theory where the $W$,
$Z^0$, the top quark and all the heavy particles have been integrated
out. In general,
$\vec C(M)$ depends on the scheme used to define the renormalized
operators. It is possible to show that $\W[\mu,M]$ can be 
written in the form 
\bea
\W[\mu,M]  = \hat M[\mu] \U[\mu, M] \hat M^{-1}[M] \, , 
 \label{monster} \eea
with
\be
\U[\mu,M]=  \left[\frac{\as (M)}{\as (\mu)}\right]^{
      \gammaz_Q / 2\beta_{ 0}} \, ,
\quad  \hat M[\mu] =
 \hat 1 +\frac{\as (\mu)}{4\pi}\J[\lambda(\mu)] \, ,
\label{mo2} \ee
where $\gammaz_Q$ is the leading order anomalous dimension matrix and 
$\J[\lambda(\mu)]$  can be obtained 
by solving the Renormalization Group Equations (RGE) at the NLO.
By defining
\be\label{eq:CRGI}
\hat{w}^{-1}[\mu] \equiv  \hat M[\mu] \left[\as (\mu) \right]^{
      - \gammaz_Q / 2\beta_{ 0}}\, , 
\ee
we get
\be
\W[\mu,M]  = \hat{w}^{-1}[\mu]\hat{w}[M]\; .  
\ee
The effective Hamiltonian~(\ref{wope}) can then be written as 
\ba\label{eq:HRGIbella}
{\cal H} 
 & = & \vec{ Q}(\mu) \cdot       \vec C(\mu) 
    =     \vec{ Q}(\mu)\hat W[\mu,M]  \vec C(M) \nonumber\\
 &   =    & \vec{ Q}(\mu) \hat{w}^{-1}[\mu] \cdot \hat{w}[M] 
                                        \vec C(M)
    =     \vec{ Q}^{RGI} \cdot \vec C^{RGI}\; ,  
\ea 
with
\be\label{eq:monster2}
\vec C^{RGI}      =  \hat{w}[M] \vec C(M)\, , \quad
\vec{Q}^{RGI}      =    \vec{ Q}(\mu) \cdot \hat{w}^{-1}[\mu]\; .
\ee
$\vec C^{RGI}$ and  $\vec{ Q}^{RGI}$ are scheme and scale independent at  
the order at which the Wilson coefficients have been computed.
\section{Comparison with other methods}
\label{sec:comparison}
In this section, we briefly discuss the relevant aspects which distinguish the lattice approach
from others which have been used in the literature to predict $\epse$. 
\par The original approach of the Munich group was to extract 
the values of the relevant matrix elements from experimental 
measurements~\cite{buras2,buras1}. This method guarantees the
consistency of the operator matrix elements  with the
corresponding Wilson coefficients.  
Unfortunately, with the Munich method it is impossible to get  
the two most important contributions, namely  those corresponding
to  $\langle Q_{6} \rangle_{0}$ and  $\langle Q_{8} \rangle_{2}$. 
 For this reason, ``guided by the results presented above
and biased to some extent by the results from the large-$N$  approach  and
lattice calculations", the authors of ref.~\cite{silvestrini} have taken
$B_6 = 1.0 \pm 0.3$ and $B_8^{(3/2)}=0.8 \pm 0.2$, and $\msms=110\pm 20$ MeV
at $\mu=1.3$ GeV. These
values, if assumed to hold in the $HV$ regularization, are very close to those used in
ref.~\cite{newnoi}. They do not come however from a calculation consistently made 
within a given theoretical approach (large $N$ expansion, $\chi$QM or lattice for example).  
\par The $1/N$ expansion and the $\chi$QM are effective theories.  To be specific, in the
framework of the $1/N$ expansion the starting point is given by 
the chiral Lagrangian for pseudoscalar mesons expanded in powers 
of masses and momenta. At the leading order in $1/N$, local  four-fermion
operators can be written in terms of products of currents and densities,
which are expressed in terms of the fields and coupling of
the effective theory.  In higher orders,  a (hard) cutoff, $\Lambda_c$, must be
introduced  to compute
the relevant loop diagrams.  The cutoff is usually identified with the scale at which 
the short-distance Wilson coefficients must be evaluated. \par 
Divergences appearing in factorizable contributions can be reabsorbed
in the renormalized coupling of the effective Lagrangian and in the
quark masses, non-factorizable corrections constitute the part
which should be matched to the short distance coefficients.
By using the  intermediate colour-singlet boson method, the authors
of refs.~\cite{pass}--\cite{paschoslast},\cite{bijnens} claim to be able to perform a
consistent matching, including the finite terms, of the matrix elements
of the operators in the effective theory to the corresponding Wilson
coefficients. It is precisely this point which, in our opinion, has never
been demonstrated in a convincing way. 
 If the  matching is ``consistent", then it should be possible
to show analytically that the cutoff dependence of the matrix elements computed in
the $1/N$ expansion cancels that of the Wilson coefficients,
at least at the order in $1/N$ at which they are working.
Moreover, if really finite terms are under control, it should be possible
to tell whether the coefficients should be taken in $HV$, $NDR$ or any
other renormalization scheme.
\par The fact that in higher orders even quadratic divergences appear,
with the result that the logarithmic divergences depend now on the
regularization, makes the matching even more problematic.
Theoretically, we cannot imagine any mechanism to cancel the cutoff
dependence of the physical amplitude in the presence of quadratic
divergences, which should, in our opinion, disappear in any reasonable
version of the effective theory.  It is also important to show (and to our knowledge it has
never been done) that the numerical
results for the matrix elements are stable with respect to  
the choice of the ultraviolet cutoff.   This would also clarify the issue of the
routing of the momenta in divergent integrals. 
% It would be surprising if the lattice
% community would make the statement that it exists only one regularized theory which gives
% the correct physical results.
For example, the matrix elements in the meson theory could be computed in some
lattice regularization.  
\section{Numerical results}
\label{sec:numerical}
As discussed above,
all the methods used in the  calculation of $\epse$ are not completely 
satisfactory and in general
suffer from large theoretical uncertainties.
\par The lattice approach can, in principle, 
compute  the relevant matrix elements without any model assumption 
(at least at the lowest order in the chiral expansion), and with operators
consistently defined to match the Wilson coefficients of the effective
Hamiltonian. In spite of these advantages the lattice results
for $\langle Q_6 \rangle_0$ are  inconclusive, as discussed before.  
Regarding the surprising result of ref.~\cite{newsoni}, 
we think that further scrutiny and confirmation from other
calculations are needed  before using it in a phenomenological analysis.  
\par In the absence of any definite result for $\langle
Q_6 \rangle_0$ from the lattice, ref.~\cite{newnoi}  assumed for 
this matrix element the value  
\be \langle Q_{6} \rangle_{0}\equiv\langle \pi \pi \vert Q^{HV}_{6} \vert 
K^{0}\rangle_{I=0} = -0.4 \pm 0.4  \, {\rm GeV}^{3}\, , \label{eq:q6hv}\ee
and 
\be\langle Q_5 \rangle_0 = 1/3 \langle Q_6 \rangle_0 \, ,\label{eq:q5} \ee
at a scale $\mu=2$ GeV. The value of the matrix element 
in eq.~(\ref{eq:q6hv}) corresponds to  
$B_{6}=1.0 \pm 1.0$  for  a ``conventional'' mass  fixed to 
$m_s^{\overline{MS}}+m_d^{\overline{MS}}=130$ MeV. 
\par For $\langle Q_{7,8} \rangle_2$,  the values of ref.~\cite{giusti} (obtained with an 
improved action using non-perturbatively  renormalized operators at $\mu=2$ GeV) 
have been used, namely
\be \langle Q_{7} \rangle_{2}\equiv \langle \pi \pi \vert Q^{HV}_{7} \vert 
K^{0}\rangle_{I=2} = 0.18\pm 0.06 \, {\rm GeV}^{3} \, ,\label{eq:q7hv}\ee
\be \langle Q_{8} \rangle_{2}\equiv \langle \pi \pi \vert Q^{HV}_{8} \vert 
K^{0}\rangle_{I=2} = 0.62 \pm 0.12 \, {\rm GeV}^{3} \, ,\label{eq:q8hv}\ee
where the superscript $HV$ denotes the t'Hooft-Veltman 
renormalization scheme.
\par By
varying the input parameters  as  described in ref.~\cite{newnoi} and
by weighting the Monte Carlo events with the experimental constraints, the 
 prediction for $\epse$ is
\be \epse= (3.6^{+6.7}_{-6.3} \pm 0.5)\times 10^{-4} \label{eq:mainresult}\, , \ee
where the third error on $\epse$ is an estimate of the residual scheme dependence
due to unknown higher-order corrections in the perturbative expansion~\footnote{
The value in eq.~\protect\ref{eq:mainresult} is slightly different from that presented
at the the Conference and quoted by ref.\protect\cite{burask99}. 
The reason is that the final analysis of ref.~\cite{giusti} found 
for $\langle Q_8\rangle_{I=2}$ a value  larger by about $15\%$ 
than the preliminary one.}.
Given the large theoretical uncertainties,  and taking into account some  differences in the 
calculation of this quantity (choice of the renormalization scale, values of several $B$ parameters, 
etc.), the  result in eq.~(\ref{eq:mainresult}) is in substantial agreement, 
though slightly lower,  with
the recently upgraded  evaluation of ref.~\cite{silvestrini}: $\epse= (7.7^{+6.0}_{-3.5}  )\times 
10^{-4}$ and $\epse= (5.2^{+4.6}_{-2.7}  )\times 10^{-4}$ in $NDR$ 
and  in  $HV$  respectively. It is also 
very close to previous  estimates of the 
Rome~\cite{ciuc1,ciuc3} and Munich 
group~\cite{buras2,buras1}. This agreement it is not surprising since the
two groups use very similar inputs for the matrix elements and 
the experimental parameters have only slightly changed in the last few
years.   The crucial question, namely a   quantitative
determination of $\langle Q_6\rangle_0$, remains unfortunately  still
unsolved. \par 
All the above   results are, however, much 
lower than the recent measurements of KTeV, $Re(\epse)=(28.0 \pm 4.1)\times 
10^{-4}$, of NA48, $Re(\epse)=(18.5 \pm 7.3)\times 10^{-4}$, or than the 
present world average $Re(\epse)_{{WA}}=(21.2 \pm 4.6)\times
10^{-4}$, 
determined by the results of refs.~\cite{KTeVK99}--\cite{NA31},\cite{E731}
\par
By scanning various input parameters (in the conventional approach 
$B_{6}$, $B_{8}^{(3/2)}$,  $\asz$, $Im\lambda_t$ etc.)
and in particular by  choosing them close to their extreme values it is  
possible to obtain   $\epse \sim 20 \times 10^{-4} $.
This also  gives  the impression of  a better agreement
(lesser disagreement) between the theoretical predictions and the data.  
For example, in ref.~\cite{newnoi} the scanning gives
$  -11 \times 10^{-4}  \le \epse \le  27 \times 10^{-4}$.
In spite of the fact that the experimental world average  is compatible
with  the ``scanned" range above,  a conspiracy of several inputs   in the same
direction is necessary in order to get  a large
value of $\epse$. For central values of the parameters, the
predictions are, in general,  much lower than the experimental results.
For this reason, barring the possibility  of new physics effects~\cite{murayama}, we
believe  that  an important message is arriving from the  experimental
results: \vskip 0.01 cm
{\it penguin contractions (or eye-diagrams, not to be confused with penguin 
operators), which are usually neglected within factorization,  
give  contributions which makes the matrix elements definitely larger than  
their factorized  values.}  \vskip 0.01 cm
This implies that the ``effective''  $B$ parameters of the relevant  operators,  
specifically  those relative to the matrix elements of $Q_{1}$ and $Q_{2}$ for 
$Re(A_{0})$ and of $Q_{6}$ for $\epse$ are much  larger than 1. This 
interpretation would provide a unique dynamical mechanism to explain   both
the  $\Delta I=1/2$ rule and   a large  value of $\epse$~\cite{ciuck99}.
Large contributions from penguin contractions  are  actually found 
by calculations performed in the framework of the Chiral Quark 
Model ($\chi$QM)~\cite{bert1}--\cite{bert2} or the $1/N$ 
expansion~\cite{pass,ham,paschoslast,1oN}.  It is very important that
these indications  find  quantitative confirmation in other approaches, 
 for example in   lattice QCD calculations.  Note that  na\"{\i}ve 
explanations of the large value of $\epse$, such as a very low value of 
$\msms$, would leave the $\Delta I=1/2$ rule unexplained. 
\par
Finally, one may try to quantify the amount of enhancement required for the
matrix element 
of $Q_{6}$ in order to explain the experimental value of $\epse$. 
A fit of $\langle Q_{6} \rangle_{0}$ to  
$Re(\epse)_{{WA}}$
gives $\langle Q_{6} \rangle_{0}=-1.2^{+0.25}_{-0.21}\pm 0.15$ GeV$^{3}$, about 
$2.5$  times larger than the central value used in our analysis.


\begin{thebibliography}{88}
\bibitem{alta} G. Altarelli, G. Curci, G. Martinelli and S. Petrarca, Nucl.~Phys. { B187} (1981) 461.
\bibitem{bur1} A.J. Buras, P.H. Weisz, Nucl.~Phys. { B333} (1990) 66.
\bibitem{bur2} A.J.~Buras, M.~Jamin, M.E~Lautenbacher and P.H. Weisz,
Nucl.~Phys.  B370 (1992) 69, Addendum, ibid. Nucl.~Phys. B375 (1992) 501.
\bibitem{buras2} A. Buras, M. Jamin and  M.E. Lautenbacher, \pl  B389 (1996) 
749.
\bibitem{bur3} A.J. Buras, M. Jamin and M.E. Lautenbacher, Nucl. Phys.  
B400 (1993) 37 and   B400 (1993) 75. 
\bibitem{buras1} A. Buras, M. Jamin and  M.E. Lautenbacher, Nucl.~Phys. B408 
(1993) 209.
\bibitem{noi} M. Ciuchini, E. Franco, G. Martinelli and L. Reina,  Nucl.~Phys.  B415 (1994) 403.
\bibitem{ciuc2} M. Ciuchini {\em et al.},  Z. Phys. C68 (1995) 239.
\bibitem{bert1} S.~Bertolini, J.O. Eeg and M.~Fabbrichesi, \np   B476  (1996) 225.
\bibitem{bert2} S.~Bertolini, J.O. Eeg, M.~Fabbrichesi and E.I.~Lashin, 
\np  B514 (1998) 93. 
\bibitem{pass}  J.~Heinrich, E.A.~Paschos, J.M.~Schwarz and Y.L.~Wu, , 
\pl B279 (1992) 140; E.A.~Paschos, review presented at the 27th 
Lepton-Photon Symposium, Beijing, China (1995).
\bibitem{1oN2} T.~Hambye {\em et al.}, \prd D58 (1998) 014017. 
\bibitem{ham} T.~Hambye, G.O.~K\"{o}hler and P.H.~Soldan,  
hep-ph/9902334.
\bibitem{paschoslast} T.~Hambye, G.O.~K\"ohler, E.A.~Paschos and 
P.H.~Soldan, hep-ph/9906434.
\bibitem{burask99} A.~Buras, talk given  this Conference
to appear in the Proceedings, hep-ph/9908395.
\bibitem{giusti} A.~Donini, V.~Gim\'enez, L.~Giusti and G.~Martinelli, BUHEP-99-23; 
L.~Giusti, presented at Latt99, June 29--July 3 1999, Pisa,
Italy, to appear in  the Proceedings, hep-lat/9909041.
\bibitem{don} J.F.~Donoghue, E.~Golowich, B.R.~Holstein and 
J.~Trampetic, \pl  B179 (1986) 361.
\bibitem{bg} A.J.~Buras and J.M.~G\'erard, \pl B192 (1987) 156.
\bibitem{lu} M.~Lusignoli, \np B325 (1989) 33. 
\bibitem{npm} G.~Martinelli, C.~Pittori, C.T.~Sachrajda, M.~Testa and A.~Vladikas, 
Nucl. Phys.~B445~(1995)~81.
\bibitem{npm4f} G.~Martinelli {\em et al.}, Nucl.~Phys.~B445~(1995)~81; 
A.~Donini {\em et al.},  Phys.~Lett.~B360~(1996)~83;
M.~Crisafulli {\em et al.}, Phys.~Lett.~B369~(1996)~325;
L. Conti {\em et al.},  Phys.~Lett.~B421~(1998)~273;  C.~R.~Allton et al.,
 Phys. Lett.~B453 (1999) 30.
\bibitem{ciuc1} M. Ciuchini, E. Franco, G. Martinelli and L. Reina,
Phys. Lett. { B301} (1993) 263.
\bibitem{ciuc3} M.~Ciuchini, \np  (Proc. Suppl.) 59 (1997) 149.
\bibitem{newnoi} M.~Ciuchini, E.~Franco, L.~Giusti, V.~Lubicz and G.~Martinelli,
in preparation.
\bibitem{mt} L.~Maiani and M.~Testa, Phys. Lett. B245 (1990) 585.
\bibitem{cfms} M.~Ciuchini, E.~Franco, G.~Martinelli, and L.~Silvestrini,
Phys.~Lett. B380 (1996) 353.
\bibitem{1oN} W.A.~Bardeen, A.J.~Buras and J.M.~G\'erard, \pl B180 
(1986) 133; \np B293 (1987) 787; \pl B192 (1987) 138.
\bibitem{kgs}  G.~Kilcup, R.~Gupta and S.~Sharpe, \prd D57 (1998) 
1654.
\bibitem{bgs}  T.~Bhattacharaya, R.~Gupta and S.~Sharpe, \prd D55 
(1997) 4036.
\bibitem{lellolin} L.~Lellouch and D.~Lin, \np (Proc. Suppl.) 73 (1999)
314.
\bibitem{kilcup} G.~Kilcup, \np B (Proc.Suppl) 20 (1991) 417; 
S.~Sharpe, \np B (Proc.Suppl) 20 (1991) 429; S.~Sharpe {\em et al.},
\pl 192B (1987) 149.
\bibitem{k2} D.~Pekurovsky and G.~Kilcup, hep-lat/9709146  and 
hep-lat/9812019.
\bibitem{stagpert} S.~Sharpe and A.~Patel, hep-lat/9310004;
N.Ishizuka and Y.~Shizawa, \prd D49 (1994) 3519.
\bibitem{newsoni} T.~Blum {\em et al.}, BNL-66731, hep-lat/9908025.
\bibitem{qmasses}  R.~D.~Kenway,  Nucl. Phys. (Proc.Suppl.) 73 (1999)~16;
S. R. Sharpe,  talk given at 29th International Conference on High-Energy Physics (ICHEP 98), 
Vancouver, Canada 1998, hep-lat/9811006;
V. Lubicz,  Nucl. Phys. (Proc.Suppl.) 74 (1999) 291. 
\bibitem{sr} S.~Ryan, talk given 
at this Conference, to appear in the Proceedings, hep-ph/9908386.
\bibitem{damir} D.~Becirevic, V.~Gim\'enez, V.~Lubicz and 
G.~Martinelli, hep-lat/9909082. 
\bibitem{wittig}  J.~Garden {\em et al.}, ALPHA and UKQCD Collaboration, DESY-99-075, 
 hep-lat/9906013.
\bibitem{claude} C.~Bernard, private comunication.
\bibitem{gimenez} V.~Gim\'enez, presented at Latt99, June 29--July 3 1999, Pisa,
Italy, to appear in  the Proceedings.
\bibitem{taron} L.~Lellouch,  E.~de~Raphael and J.~Taron, Phys.~Lett. B414 (1997) 195.
\bibitem{silvestrini} S.~Bosch {\em et al.}, hep-ph/9908408.
\bibitem{bijnens} J.~Bijnens and J.~Prades, JHEP 01 (1999) 023.
\bibitem{KTeVK99} KTeV Collaboration, A.~Alavi-Harati {\em et al.},
\prd Lett. 83 (1999) 22.
\bibitem{NA48K99} NA48 Collaboration, talk given by M.~Sozzi at Kaon99, June 
21--26 1999, Chicago, USA, to appear in the Proceedings.
\bibitem{NA31}  NA31 Collaboration, H.~Burkhardt {\em et al.},
\pl B206 (1988) 169; G.D.~Barr {\em et al.}, \pl B317 (1993) 233.
\bibitem{E731} E731 Collaboration, L.K.~Gibbons {\em et al.}, \prd Lett.
70 (1993) 1203. 
\bibitem{murayama} H.~Murayama, talk given 
at this Conference, to appear in the Proceedings, hep-ph/9908442.
\bibitem{ciuck99} M.~Ciuchini {\em et al.}, talk given by M.~Ciuchini 
 at this Conference, to appear in the Proceedings.
\end{thebibliography}
\end{document}